\documentstyle[12pt,a4]{article}
\begin{document}
\begin{titlepage}
\begin{flushright}
April 1998
\end{flushright}
\vspace*{1.6cm}
\begin{center}
{\Large\bf The $\eta$-$\eta\prime$ mixing angle revisited}\\
\vspace*{1cm}
A.~Bramon\\
{\footnotesize{\em
       Departament de F{\'\i}sica,\\
       Universitat Aut\`onoma de Barcelona,
       E-08193 Bellaterra (Barcelona), Spain}}\\
\vspace*{0.4cm}
R.~Escribano\\
{\footnotesize{\em
       Service de Physique Th\'eorique,\\
       Universit\'e Libre de Bruxelles,
       CP 225, B-1050 Bruxelles, Belgium}}\\
\vspace*{0.4cm}
M.~D.~Scadron\\
{\footnotesize{\em
       Physics Department, University of Arizona,
       Tucson, AZ 85721, USA}}\\
\end{center}
\vspace*{1cm}
\begin{abstract}
The value of the $\eta$-$\eta\prime$ mixing angle $\theta_P$ is 
phenomenologically deduced from a rather exhaustive and up-to-date 
analysis of data including strong decays of tensor
and higher-spin mesons, electromagnetic decays of vector and pseudoscalar
mesons, $J/\psi$ decays into a vector and a pseudoscalar meson, and other
transitions. A value of $\theta_P$ between $-17^\circ$ and $-13^\circ$ is
consistent with the present experimental evidence and the average
$\theta_P=-15.5^\circ\pm 1.3^\circ$ seems to be favoured.
\end{abstract}
\end{titlepage}
\section{Introduction}
\label{intro}
The value of the $\eta$-$\eta\prime$ mixing angle in the pseudoscalar-meson 
nonet has been discussed many times in the last thirty years. 
Quite possibly it has become one of the most interesting $SU(3)$-breaking
hadronic parameters to measure since $SU(3)$ symmetry was proposed.
In recent years three independent analyses have surveyed world data 
indirectly measuring this angle. A well known contribution to this discussion 
is the phenomenological analysis performed by Gilman and Kauffman
\cite{GK} almost a decade ago. 
The approximate value $\theta_P\simeq -20^\circ$ 
(see Sec.~II for notation and definitions) was proposed by
these authors through a rather complete discussion of the experimental
evidence available at that time. Another analysis by two of the present
authors \cite{BS} concluded that a somewhat less negative value,
$\theta_P=-14^\circ\pm 2^\circ$, seems to be favoured. A significant
difference between  these two independent analyses concerns the set of rich
data on $J/\psi$  decays into a vector and a pseudoscalar meson, $J/\psi\to
VP$, which were included in the first analysis \cite{GK} but not in the 
second one \cite{BS}. Finally, the more recent discussion involving several 
channels performed by Ball, Fr\`ere and Tytgat \cite{BFT} has led to 
$\theta_P$ between $-20^\circ$ and $-17^\circ$.

Our purpose in the present paper is to obtain a new value of this 
$\eta$-$\eta\prime$ mixing angle along the lines of the previous works. To 
this aim, we will perform a rather exhaustive and
updated analysis using the available world data \cite{PDG} and well 
established phenomenology on strong interaction decays of meson resonances 
into pseudoscalar pairs, electromagnetic decays of low mass mesons, $J/\psi$
decays into a vector and a pseudoscalar meson, and other transitions. Our
main assumptions are the validity of $SU(3)$ symmetry and, quite often, the
stronger condition of nonet symmetry to relate the $SU(3)$-octet to the
$SU(3)$-singlet. We also introduce $SU(3)$-breaking corrections in terms of
constituent quark mass differences when their effects can be controlled 
and/or computed. In this sense, we define $\bar m\equiv (m_u+m_d)/2$ and 
take $m_s/\bar m\simeq  1.45$ from previous phenomenological analyses 
\cite{QCM}.
Finally, we also assume the $\eta$-$\eta\prime$ to form a simple two-state 
system and neglect possible mixing with other pseudoscalar states, in 
particular with glueballs; we therefore consider just one single and real 
mixing angle.

The paper is organized as follows. In Sec.~2 we introduce the notation and
interrelate quark content and mixing angles. Sections 3 and 4 cover the
strong decays into two pseudoscalars of spin-two (tensor) mesons, 
$T\rightarrow PP$, and higher-spin mesons, 
$M_J\rightarrow PP$ with $J=3,4\ldots$, respectively. Electromagnetic
radiative decays involving vector and pseudoscalar mesons, 
$V\rightarrow P\gamma$ and $P\rightarrow V\gamma$, are 
discussed in Sec.~5. In Sec.~6 we consider the two-photon annihilation decays
$\pi^0,\eta,\eta\prime\rightarrow \gamma\gamma$. Sec.~7 deals with $J/\psi$
decays into a vector and a pseudoscalar meson, $J/\psi\rightarrow VP$. Finally,
in Sec.~8 we briefly present results on other transitions and in Sec.~9 we
summarize our conclusions.
\section{Notation}
\label{notation}
Throughout this section we fix our notation which follows quite closely that 
introduced by Gilman and Kaufman \cite{GK} and previous work by 
Rosner \cite{ROS}. 
The $SU(3)$-octet and -singlet states are
\begin{equation}
\label{eta8&eta0}
|\eta_{8}\rangle=\frac{1}{\sqrt{6}}|u\bar u+d \bar d-2s \bar s\rangle\ ,\ \
|\eta_{0}\rangle=\frac{1}{\sqrt{3}}|u\bar u+d \bar d+s \bar s\rangle\ ,
\end{equation}
and, in terms of this $SU(3)$ basis, the physical $\eta$ and $\eta\prime$ 
states are defined to be 
\begin{equation}
\label{eta&etaprime}
\begin{array}{l}
|\eta\rangle=
\cos\theta_{P}|\eta_{8}\rangle-\sin\theta_{P}|\eta_{0}\rangle\ ,\\[1ex]
|\eta\prime\rangle=
\sin\theta_{P}|\eta_{8}\rangle+\cos\theta_{P}|\eta_{0}\rangle\ .
\end{array}
\end{equation}
For some purposes it is more convenient to use the so-called 
nonstrange(NS)-strange(S) quark basis:
\begin{equation}
\label{etaq&etaprimeq}
\begin{array}{l}
|\eta\rangle=
X_{\eta}\frac{1}{\sqrt{2}}|u\bar u+d \bar d\rangle+Y_{\eta}|s \bar s\rangle
\equiv \cos\varphi_{P}|\eta_{NS}\rangle-\sin\varphi_{P}|\eta_{S}\rangle\
,\\[1ex]
|\eta\prime\rangle=
X_{\eta\prime}\frac{1}{\sqrt{2}}|u\bar u+d \bar d\rangle+
Y_{\eta\prime}|s \bar s\rangle
\equiv \sin\varphi_{P}|\eta_{NS}\rangle+\cos\varphi_{P}|\eta_{S}\rangle\ ,
\end{array}
\end{equation}
where $|\eta_{NS}\rangle=|u\bar u+d\bar d\rangle/\sqrt{2}$ and
$|\eta_{S}\rangle=|s\bar s\rangle$. Assuming the orthogonality of the 
physical $\eta$-$\eta\prime$ states and no mixing with other pseudoscalars, 
one has
\begin{equation}
\label{oneangle}
X_{\eta}^{2}+Y_{\eta}^{2}=
X_{\eta\prime}^{2}+Y_{\eta\prime}^{2}=1\ ,\ \
X_{\eta}X_{\eta\prime}+Y_{\eta}Y_{\eta\prime}=0\ .
\end{equation}
In this case, just a single and real mixing angle governs the whole mixing 
phenomena if any energy dependence is (as usual) neglected. In terms of
$\theta_P$ or $\varphi_P$ the $X$'s and $Y$'s can be written
\begin{equation}
\begin{array}{l}
X_{\eta}=Y_{\eta\prime}\equiv\cos\varphi_{P}=
\frac{1}{\sqrt{3}}\cos\theta_{P}-\sqrt{\frac{2}{3}}\sin\theta_{P}\
,\\[1ex]
Y_{\eta}=-X_{\eta\prime}\equiv-\sin\varphi_{P}=
-\sqrt{\frac{2}{3}}\cos\theta_{P}-\frac{1}{\sqrt{3}}\sin\theta_{P}\ ,
\end{array}
\end{equation}
with $\theta_{P}=\varphi_{P}-\arctan\sqrt{2}\simeq\varphi_{P}-54.7^\circ$ and,
conversely,
\begin{equation}
\tan\theta_{P}=
-\frac{\sqrt{2}X_{\eta}+Y_{\eta}}{X_{\eta}-\sqrt{2}Y_{\eta}}=
\frac{X_{\eta\prime}-\sqrt{2}Y_{\eta\prime}}
     {\sqrt{2}X_{\eta\prime}+Y_{\eta\prime}}\ .
\end{equation}

In most of the next sections, we start with the presentation of the 
phenomenological and SU(3)-symmetric lagrangians responsible for the different
transitions. Then SU(3)-breaking effects controlled by constituent quark 
mass differences are introduced when their origin is understood and their 
effects can be computed. 
A common feature of these lagrangians is the appearance of the $SU(3)$ matrix 
$P$ containing the fields of the pseudoscalar meson nonet and their 
derivatives. 
The normalization of the SU(3) matrix $P$ is such that its diagonal elements 
are
$\pi^0/{\sqrt 2}+\eta_8/{\sqrt 6}+\eta_0/{\sqrt 3}$,
$-\pi^0/{\sqrt 2}+\eta_8/{\sqrt6}+\eta_0/{\sqrt 3}$ and
$-2\eta_8/{\sqrt6}+\eta_0/{\sqrt3}$. 
Similar $SU(3)$ matrices $V^\mu$, $T^{\mu\nu}\ldots$ are introduced for the
nonets of vector, tensor and higher-spin mesons, and mixing phenomena inside 
these nonets are consistently taken into account. Physical amplitudes are 
extracted and the corresponding theoretical decay widths are computed and 
compared with the available data. As a result of the corresponding fits, 
independent estimates of the $\eta$-$\eta\prime$ mixing  angle are obtained 
and discussed in each section.
\section{Strong Decays of Tensor Mesons $T(2^{++})\rightarrow PP$}
The phenomenological and SU(3)-symmetric lagrangian for these 
$T\rightarrow PP$ decays is
\begin{equation}
{\cal L}_{TPP}=
g\,tr(T^{\mu\nu}\{P,\partial_{\mu}\partial_{\nu}P\}_+)=
g\,tr(T^{\mu\nu}(P\partial_{\mu}\partial_{\nu}P+
(\partial_{\mu}\partial_{\nu}P)P))\ ,
\end{equation}
where $P$ and $T^{\mu\nu}$ are the $SU(3)$-nonet matrices mentioned in the
previous section and $g$ is a generic strong-interaction coupling constant. 
Similarly, we define the $f$-$f\prime$ mixing angle in this 
tensor-meson nonet ($J^{PC}=2^{++}$) in a way analogous to the pseudoscalar 
case (see Ref.~\cite{PDG}):
\begin{equation}
\label{f&fprime}
\begin{array}{l}
|f\rangle=
\cos\varphi_{T}|f_{NS}\rangle-\sin\varphi_{T}|f_{S}\rangle\ ,\\[1ex]
|f\prime\rangle=
\sin\varphi_{T}|f_{NS}\rangle+\cos\varphi_{T}|f_{S}\rangle\ ,
\end{array}
\end{equation}
with
$\varphi_{T}\equiv\varphi_{2}=\theta^{PDG}_{T}-\arctan{1/\sqrt{2}}\simeq
28^{\circ}-35.3^{\circ}=-7.3^{\circ}$.
This small value for the mixing angle follows from the quadratic 
Gell-Mann--Okubo (GMO) mass formula \cite{PDG} thus implying an almost ideal 
mixing in the tensor-meson nonet.

In Table \ref{tableTPP} we present for each strong tensor-meson decay
both the normalized coupling of the process and the experimental branching
ratio. It is straightforward to obtain the theoretical decay amplitude and 
partial width
\begin{equation}
\label{TPP}
\Gamma(T\rightarrow PP)=
\frac{g^{2}_{TPP}}{60\pi}\frac{|\vec p_{P}|^{5}}{m_{T}^{2}}\ ,
\end{equation}
where $g_{TPP}$ is defined in Table \ref{tableTPP}, $\vec p_{P}$ is the 
momentum of the outgoing pseudoscalar meson and $m_T$ is the mass of the 
decaying tensor resonance. The symmetry factor in case two identical 
pseudoscalar mesons were produced is included in the couplings.
The couplings are assumed to be $SU(3)$ symmetric since in these decays one is
not able to control the $SU(3)$ breaking corrections via $\bar m/m_s$.

Comparing the theoretical decay widths with the experimental data
taken from \cite{PDG} (see Table \ref{tableTPP}), we extract four independent
determinations of the mixing angle $\varphi_P$. Each determination is based on
a fit performed with the same initial tensor resonance $T$ decaying into 
different $PP$ channels. 
In every case, the quality of the fits is very good and the
errors in $\varphi_P$ ---coming only from the experimental error in the 
branching ratio (BR) but not from that on the total width of the
decaying resonance--- are quite small. These four independent 
determinations of $\varphi_P$ are fully consistent. However two warnings are
worthwhile: first, although the two values obtained for $\varphi_T$ in Table
\ref{tableTPP} ($\varphi_T= -7.8^\circ\pm 2.6^\circ$ and
$\varphi_T= -2.3^\circ\pm 0.2^\circ$) reasonably agree with the approximate 
value $\varphi_T\simeq -7.3^\circ$ coming from the Gell-Mann--Okubo mass 
formula, they are slightly diverging; second, when trying to fit the 
experimental value $g_{f\rightarrow K\bar K}/g_{a_2\rightarrow K\bar K}=
1.51 \pm 0.15$ with the theoretically predicted ratio in the good SU(3) 
limit $g_{f\rightarrow K\bar K}/g_{a_2\rightarrow K\bar K}=
\cos\varphi_{T}-\sqrt{2}\sin\varphi_{T}$, one gets 
$\varphi_{T}={-25^\circ}^{+8^\circ}_{-11^\circ}$ a too negative value. 

A global fit involving all the measured $T\rightarrow PP$ decays and 
thus requiring the corresponding experimental total width of the decaying 
tensor mesons has also been performed. It leads to 
$\varphi_{P}=44.2^\circ\pm 1.4^\circ$ (or
$\theta_{P}=-10.5^\circ\pm 1.4^\circ$) and to
$\varphi_{T}\equiv\varphi_{2}=-2.9^\circ\pm 0.3^\circ$. The quality of this
global fit is much poorer (the $\chi^2$ per degree of freedom is 
$\chi^2/d.o.f=6.2$) 
than the previous partial fits as a consequence of the two warnings just
mentioned. 

In spite of this, one can conclude that the available data on strong 
$T\rightarrow PP$ decays seem to favour the value for the pseudoscalar  
mixing angle $\varphi_P \simeq 42^\circ$ (or $\theta_P
\simeq -13^\circ$).  This result confirms the conclusions presented in 
\cite{GK,BS}. The $T\rightarrow PP$ decays were not considered in \cite{BFT}.
\section{Other Strong Decays $M_J\rightarrow PP$, $J>2$}
In this section we discuss the strong interaction decays into pseudoscalar 
pairs of meson resonances with spin $J$ higher than two, $M_J\rightarrow PP$. 
Following the standard nomenclature, these resonances belong either to the 
``normal'' spin-parity 
series with $P=(-)^J$ or to the ``abnormal'' one. In the first case, one has 
$J^{PC}=4^{++}, 6^{++}\ldots$ and the situation is similar to the $2^{++}$ 
case already discussed; in the second one, with $J^{PC}=3^{--}, 5^{--}\ldots$ 
the similarities are with the well-known case of vector mesons, 
$1^{--}$. The phenomenological lagrangian needed for both series of 
higher-spin meson decays is 
\begin{equation}
\label{lagMJPP}
{\cal L}_{M_JPP}=-i^Jg\, 
tr(T^{\mu_1\mu_2\ldots \mu_J}\{P,\partial_{\mu_1}\partial_{\mu_2}\cdots
\partial_{\mu_J}P\}_\pm)\ ,
\end{equation}
where $\{P,\partial_{\mu_1}\partial_{\mu_2}\cdots \partial_{\mu_J}P\}_\pm$
stands for the anticommutator in case of even spin $J$ (positive $C$) 
or commutator in case of odd spin $J$ (negative $C$), 
as required by charge conjugation invariance. The previous lagrangians are 
taken to be nonet-symmetric because $SU(3)$-breaking effects
linked to quark mass differences cannot be controlled. In Table
\ref{tableMJPP} we show the coupling constants $g_{JPP}$ for the decay 
processes we are interested in. It is  then straightforward to calculate the 
theoretical decay rate
\begin{equation}
\label{MJPP}
\Gamma(M_J\rightarrow PP)=
\frac{g^{2}_{JPP}}{4\pi}\frac{J!}{2(2J+1)!!}
\frac{|\vec p_{P}|^{2J+1}}{m^2_J}\ .
\end{equation}

Concerning the experimental input, data on these high-spin mesons are 
rather scarce. In two cases, however, they
can be useful to extract new values for the mixing angle $\varphi_P$. Indeed, 
the three measured branching ratios for $f_4(2044)$ lead to the values
$\varphi_{P}=41.2^\circ\pm 3.7^\circ$ (or $\theta_{P}=-13.5^\circ\pm
3.7^\circ$) and $\varphi_{4}=15.7^\circ\pm 4.4^\circ$ shown in Table
\ref{tableMJPP} ($\varphi_{4}$ is defined as the mixing angle of the system
$f_4(2044)$-$f_4(2220)$). Independently, the value  
$\varphi_P=50^\circ\pm 26^\circ$ 
can be obtained from the two measured branching
ratios of $K^\ast_3(1780)$. 
Notice that in this case one has a much larger error
even if one started with the rather accurately measured branching ratio 
$BR(K^\ast_3\rightarrow K\eta/K^\ast_3\rightarrow K\pi)_
{\mbox{\footnotesize exp}}=0.41\pm 0.08$. 
This is due to the fact that the theoretical ratio 
$BR(K_J^\ast \rightarrow K\eta/K_J^\ast \rightarrow K\pi)=
1/3(\cos\varphi_P+(-)^{J+1}\sqrt{2}\sin\varphi_P)^2
(\vec p_{\eta}/\vec p_{\pi})^{2J+1}$ contains the sign
$(-)^{J+1}$ due to charge conjugation invariance. For the actual values 
of $\varphi_P$, this sign makes the
dependence of this ratio on $\varphi_P$ rather smooth for $J$ odd, as we have 
just seen. 
On the contrary, that dependence is much stronger for $J$ even, but then the
ratio has to be very small and no data are known except for the case of 
$K^\ast_2$, as discussed in the previous section. 

As a conclusion for this section, we can say that a pseudoscalar-mixing angle
of $\varphi_{P}\simeq 41^\circ$ (or $\theta_{P}\simeq -14^\circ$) is favoured 
again from our simple SU(3) analysis of  $M_J\rightarrow PP$, $J>2$, decays 
and particularly from those of $f_4$(2050). 
This is a new result since these $M_J\rightarrow PP$ decays were not 
considered in previous analyses.
\section{Radiative Decays $V\rightarrow P\gamma$, $P\rightarrow V\gamma$}
We start this section with the phenomenological lagrangian that conventionally
accounts for the amplitudes of the decay processes $V\rightarrow P\gamma$ and 
$P\rightarrow V\gamma$
\begin{equation}
\label{lagVPg}
{\cal L}_{VP\gamma}=
g\,\epsilon_{\mu\nu\alpha\beta}\,\partial^{\mu}A^{\nu}\,
tr(Q(\partial^{\alpha}V^{\beta}P+P\partial^{\alpha}V^{\beta}))\ ,
\end{equation}
where $g$ is a generic, electromagnetic coupling constant,
$\epsilon_{\mu\nu\alpha\beta}$ is the totally antisymmetric Levi-Civita 
tensor, $A_\mu$ is the photon field, $P$ is the pseudoscalar meson matrix, 
$V_\mu$ its vector counterpart and $Q$ is the quark-charge matrix
$Q=diag\{2/3,-1/3,-1/3\}$. From the previous lagrangian, it is easy to
calculate the theoretical decay widths
\begin{equation}
\label{gammaVPg}
\Gamma(V\rightarrow P\gamma)=
\frac{1}{3}\frac{g^{2}_{VP\gamma}}{4\pi}|\vec p_{\gamma}|^{3}=
\frac{1}{3}\Gamma(P\rightarrow V\gamma)\ ,
\end{equation}
where $g_{VP\gamma}$ is the specific coupling constant for each process
defined in Table \ref{tableVPg&PVg} and $|\vec p_{\gamma}|$ is the momentum of
the final photon. We have computed all these transition amplitudes in the
framework of the quark model with $SU(3)$ and nonet symmetry broken 
by constituent quark mass differences according to a well known and 
time-honored prescription. It amounts to a modification in the original 
charge quark matrix $Q$ 
via the introduction of the multiplicative $SU(3)$-breaking term 
$1-s_e\equiv\bar m/m_s\simeq 1/1.45$ in the $s$-quark charge entry, 
as required in these magnetic-dipolar transitions if one takes into account 
the well known differences between the light- and strange-quark magnetic 
moments. 
Contrasting with the two preceeding sections, in the present case
we can easily control and compute the effects of these corrections. Moreover,
in our analysis, the apparently negligible effects of non-ideal mixing in the 
vector-meson nonet will be taken into account. Indeed, we introduce the 
small, but certainly non-vanishing, departure of $\omega$ and $\phi$ from the 
ideally mixed states
$\omega_{NS}\equiv (u\bar u+d\bar d)/\sqrt{2}$ and $\phi_{S}\equiv s\bar s$ 
by writing the physical states in the nonstrange-strange basis as
\begin{equation}
\begin{array}{l}
\label{omega&phi}
|\omega\rangle=
\cos\varphi_{V}|\omega_{NS}\rangle-\sin\varphi_{V}|\phi_{S}\rangle\ ,\\[1ex]
|\phi\rangle=
\sin\varphi_{V}|\omega_{NS}\rangle+\cos\varphi_{V}|\phi_{S}\rangle\ ,
\end{array}
\end{equation}
where $\varphi_{V}$ is a small angle signalling departure from ideal mixing.
The absolute value and relative sign of the $\omega$-$\phi$ mixing angle are 
well known,  
$\sin\varphi_{V}\simeq\tan \varphi_{V}=+0.059\pm 0.004$ or $\varphi_{V}\simeq
+3.4^\circ$, and come from the clearly understood ratio \cite{PDG,JS}
$\Gamma (\phi\to \pi^0\gamma)/\Gamma(\omega\to\pi^0\gamma) = 
\tan^2 \phi_V (p_\phi / p_\omega)^3 = (8.10\pm 0.94)\times 10^{-3}$ and the 
$\omega$-$\phi$ interference effects measured in
$e^+ e^- \rightarrow \pi^+\pi^-\pi^0$ annihilation data \cite{10PL,BGP}. 
However, in our analysis we have not fixed this angle to the above value but 
has been left as a free parameter to fit. 

Table \ref{tableVPg&PVg} displays all the decay channels involved in 
our discussion together with their theoretical amplitudes extracted
from the lagrangian (\ref{lagVPg}), as well as the experimental values for the 
respective decay widths taken from \cite{PDG}. We have performed a global fit 
to all these decay widths in order to find out the most suitable 
$\eta$-$\eta\prime$ mixing angle.
In addition, a fitted value of the $\omega$-$\phi$ mixing angle is also 
obtained.
The fit is excellent ($\chi^2/d.o.f=1.4$) and the data seems to prefer 
the values $\varphi_{P}=36.5^\circ\pm 1.4^\circ$ 
(or $\theta_{P}=-18.2^\circ\pm 1.4^\circ$)
and $\varphi_{V}=3.4^\circ\pm 0.2^\circ$. This value of $\varphi_{P}$ nicely 
agrees with the ones proposed by Gilman and Kauffman \cite{GK} and by 
Ball {\it et al.} \cite{BFT}, but it is somewhat smaller than the one 
favoured in \cite{BS}. Concerning the value of $\varphi_{V}$ it perfectly 
agrees with the one coming from the well known  Gell-Mann--Okubo mass
formula ($\varphi_{V}\simeq 39^\circ - 35.3^\circ=+3.7^\circ$, see \cite{PDG})
and (including the sign) with the previously mentioned values coming from 
radiative  $\omega$ and $\phi$ decays and $\omega$-$\phi$ interference in
$e^+e^-\rightarrow \pi^+\pi^-\pi^0$ \cite{10PL,BGP}. This agreement represents
an important test of the correctness of our treatment. 

Another, more crucial test, originally proposed by Rosner \cite{ROS} 
and expected to be measured at DA$\Phi$NE $\phi$-factory in the near future, 
to elucidate the definite value for $\varphi_{P}$ is the measurement of the 
ratio
\begin{equation}
\label{test}
R_\phi\equiv
\frac{\Gamma(\phi\rightarrow\eta\prime\gamma)}
     {\Gamma(\phi\rightarrow\eta\gamma)}=
\cot\varphi_{P}^{2}(1-\frac{m_{s}}{\bar m}
\frac{\tan\varphi_{V}}{\sin 2\varphi_{P}})^{2}
\left(\frac{p_{\eta\prime}}{p_{\eta}}\right)^{3}\ .
\end{equation}
This ratio predicts $7.6\times 10^{-3}$ for $\varphi_{P}=35^{\circ}$
$(\theta_{P}\simeq-20^{\circ})$ and $5.6\times 10^{-3}$ for 
$\varphi_{P}=39.2^{\circ}$ $(\theta_{P}=-15.5^{\circ})$,
well within the expected capabilities of DA$\Phi$NE.
A recent experimental measurement \cite{AKH} of the branching ratio
$BR(\phi\rightarrow\eta\prime\gamma)=1.2^{+0.7}_{-0.5}\cdot 10^{-4}$
yields $R_\phi=9.5^{+5.2}_{-4.0}\cdot 10^{-3}$, with an error still too 
large to decide between the previous predicted values.
\section{$P^0\rightarrow\gamma\gamma$}
We begin the discussion giving the well known phenomenological lagrangian
\begin{equation}
\label{lagPgg}
{\cal L}_{P^{0}\gamma\gamma}=
g\,\epsilon_{\mu\nu\alpha\beta}\,
\partial^{\mu}A^{\nu}\partial^{\alpha}A^{\beta}\,tr(Q^{2}P)\ ,
\end{equation}
which describes the annihilation of a neutral pseudoscalar meson $P^{0}$ 
into two photons. 
In a straightforward manner one can extract from the previous lagrangian the
theoretical decay rate for the various $P^0\rightarrow \gamma\gamma$ processes
\begin{equation}
\label{gammaPgg}
\Gamma(P^0\rightarrow \gamma\gamma)=
g^{2}_{P\gamma\gamma}\frac{1}{64\pi}m_{P}^{3}\ ,
\end{equation}
where $g_{P\gamma\gamma}$ is the coupling constant for each process
presented in Table \ref{tablePgg} and $m_{P}$ is the mass of the decaying 
pseudoscalar meson. As in the previous section, $SU(3)$-breaking effects 
driven by the constituent quark mass ratio $\bar m/m_s$ can be controlled 
since they appear through a modification in the quark charge matrix $Q$ 
similar to the previous case. For $\bar m/m_s\simeq 1/1.45$, a comparison 
of the theoretical decay rates of the processes
$\pi^0\rightarrow\gamma\gamma$, $\eta\rightarrow\gamma\gamma$ and
$\eta\prime\rightarrow\gamma\gamma$ with their experimental values is
presented in Table \ref{tablePgg}. The result of the global fit leads to
$\varphi_{P}=42.4^\circ\pm 2.0^\circ$ 
(or $\theta_{P}=-12.3^\circ\pm 2.0^\circ$).

The quality of the fit is again reasonably good ($\chi^2/d.o.f.=2.8$). 
The value for $\varphi_{P}$ presented here agrees with that obtained in 
\cite{BS} when quark-mass corrections were taken into account. However, our 
present value slightly disagrees with the one in \cite{GK}, 
the main reason being the discrepancy existing in the ratio 
$\Gamma(\eta\rightarrow \gamma\gamma)/\Gamma(\pi^{0}\rightarrow \gamma\gamma)$
used in \cite{GK} and its updated value (see Ref.~\cite{PDG}) used in our 
present discussion which is nearly $20\%$ smaller. The independent analysis 
by Pham \cite{PHA} for these processes lead to 
$\theta_{P}=-18.4^{\circ}\pm 2.0^\circ$.

In principle, these $P\rightarrow\gamma\gamma$ decay modes could also be
studied in the context of Chiral Perturbation Theory (ChPT) as did, for 
instance, in Ref.~\cite{DOBLE}. Then one has to face the problem of the
non-Goldstone nature of the high mass $\eta\prime$-meson. Recent attempts
along these lines lead to $\theta_{P}=-22.0^{\circ}\pm 3.3^\circ$
\cite{VEN} or to the need to consider two mixing angles of
$\simeq-20^{\circ}$ and $\simeq-4^{\circ}$ \cite{LEU1}.
ChPT could also predict $\theta_{P}$ by means of pseudoscalar masses,
but the situation is unclear as mentioned in \cite{BES} and described in 
much more detail by Leutwyler (Ref.~\cite{LEU2}).
\section{$J/\psi$ Decays}
Here we discuss the value for the $\eta$-$\eta\prime$ mixing angle that one 
can extract from the analyses of $J/\psi$ decays into a vector plus a 
pseudoscalar, $J/\psi\rightarrow VP$. Previous studies of this subject 
have appeared in the literature for the last ten  years. 
A first exhaustive analysis performed by the Mark III Collaboration 
\cite{COF} on the decays of $J/\psi$ into $VP$ concluded that the $\eta$ and
$\eta\prime$ were both consistent in being composed only of up, down and 
strange quarks and yield to a value of $\theta_P=-19.2^\circ\pm 1.4^\circ$ 
for the pseudoscalar mixing angle. Another equally  
exhaustive analysis performed by the DM2 Collaboration \cite{JOU} on the same
$J/\psi \rightarrow VP$ decays reaches similar conclusions: the $\eta$
and $\eta\prime$ mesons are consistent with a pure $q\bar q$ structure and a 
value for the mixing angle of $\theta_P=-19.1^\circ\pm 1.4^\circ$ is obtained.
Using only the data for the $J/\psi$ decays into $VP$, Morosita {\it et al.}
\cite{JAP} obtained a value of $\theta_P=-20.2^\circ$, but a more extensive 
analysis
by the same authors including also the $J/\psi$ decays into $P\gamma$ leads
to the value  $\theta_P=-18.3^\circ$. Finally, a value of $\theta_P\sim
-19^\circ$ and the conclusion that gluonium contaminations do not seem to be
present or, at least, are not required in the $\eta$-$\eta\prime$ system was 
similarly defended in \cite{AB}. In summary, all these analyses 
unanimously favour a value of $\theta_P\simeq -19^\circ$. 
A very recent work \cite{BES} performed by the present authors dealing with 
the same relevant set of  $J/\psi\rightarrow VP$ decay data
leads, however, to a value of $\theta_P=-16.9^\circ \pm 1.7^\circ$. This last 
analysis follows quite closely  the just mentioned analyses in 
Refs.~\cite{COF,JOU,JAP} except that the apparently negligible effects of 
non-ideal mixing in the vector-meson nonet, which turn out to be important, 
are fully taken into account in Ref.~\cite{BES}

In this section we do not intend to repeat the complete and exhaustive 
analysis on $J/\psi$ decays into $VP$ performed in Ref.~\cite{BES} 
but simply quote the main features and results. The relevant 
theoretical amplitudes and their corresponding experimental branching
ratios can be seen in Table \ref{tableJVP}. The origin of the various terms 
in the different amplitudes and the definitions for the parameters involved 
are explained in detail in Ref.~\cite{BES} 
but are essentially the same in all the previously mentioned earlier 
analyses.  
As stated before, however, our amplitudes in Table \ref{tableJVP} refer to 
the unmixed states $\omega_{NS}$ and $\phi_S$ rather than to the physical, 
mixed states. 
The required physical amplitudes have to be obtained by means of 
Eq.~(\ref{omega&phi}). 
The three decay modes in the upper part of the table represent 
isospin-violating 
transitions between an isoscalar initial state and an isovector final one; 
they are driven by a common isospin-violating, electromagnetic amplitude $e$ 
times a factor accounting for the quarks involved in each transition. 
The second part of the table lists transitions proceeding 
both through the isospin-violating amplitude $e$ and to the isospin-conserving
strong amplitudes $g$ and $rg$ associated with connected and disconnected 
gluonic diagrams, respectively (see Refs.~\cite{COF,JOU} for details). Also 
SU(3)-breaking is taken into account through the parameters $x$ and 
$s_e=1-\bar{m}/m_s$ (see Ref.~\cite{BES}).

The main results of this analysis, already presented in \cite{BES},
are the following: 

{\it i)} using a simple and widely accepted model, an excellent 
partial fit  ($\chi^2/d.o.f.$ = 0.7) of the decays involving a final state 
with isospin $I=1$
($J/\psi\rightarrow\rho\eta$, $\rho\eta\prime$, $\omega\pi^0$) leads to a 
value of $\varphi_{P}=40.2^\circ\pm 2.8^\circ$ 
(or $\theta_{P}=-14.5^\circ\pm 2.8^\circ$). 

{\it ii)} a global fit to all the decay modes using a more
sophisticated but also widely accepted model \cite{BES} leads similarly to
$\varphi_{P}=37.8^\circ\pm 1.7^\circ$ (or $\theta_{P}=-16.9^\circ\pm
1.7^\circ$). One also obtains an excellent value for the $\omega$-$\phi$ 
mixing angle $\varphi_V=+3.5^\circ\pm 2.2^\circ$.

A value of $\varphi_{P}\simeq 39^\circ$ 
seems to be favoured again by the cleanest subset of experimental data
involving $I=1$ final states. The global set of data (now including also 
$I=0$ final states) is affected by smaller error bars and 
seems to confirm the same result although a much more 
complicated description is needed. As a conclusion, we can say that the whole
analysis performed here improves previous analyses thanks to the
introduction of a non-negligible $\omega$-$\phi$ mixing angle $\varphi_{V}$, 
whose correct value is consistently reproduced when performing the fits.
The value we have obtained, $\theta_P \simeq - 16^\circ$,
is clearly favoured over those coming from the earlier analyses, 
$\theta_P \simeq -19^\circ$.  
\section{Other transitions}
In this section we discuss other processes related to the
$\eta$-$\eta\prime$ mixing angle which have been considered by several 
authors.
A classical example is the ratio between the
reactions $\pi^-p\rightarrow\eta n$ and $\pi^-p\rightarrow\eta\prime n$
\cite{GK}. At very high energies the difference in phase space for the two
processes becomes negligible and nonet-symmetry predicts the ratio of cross
sections
\begin{equation}
\frac{\sigma(\pi^-p\rightarrow\eta\prime n)}{\sigma(\pi^-p\rightarrow\eta n)}=
\tan^2\varphi_P\ .
\end{equation}
There exist some discrepancy concerning the experimental value of this ratio.
For completeness, we quote the two early results already considered 
in \cite{GK}. 
One result \cite{APE} leads to $\varphi_{P}=36.7^\circ\pm 1.4^\circ$ (or 
$\theta_{P}=-18.0^\circ\pm 1.4^\circ$) while the other \cite{STA} leads to
$\varphi_{P}=39.7^\circ\pm 1.0^\circ$ (or $\theta_{P}=-15^\circ\pm 1^\circ$).
More recently, a dedicated analysis by the Crystal Barrel Collaboration
\cite{AMS} favors a mixing angle of 
$\varphi_{P}=37.4^\circ\pm 1.8^\circ$ 
(or $\theta_{P}=-17.3^\circ\pm 1.8^\circ$).

Independent information comes from the
recent analysis of semileptonic $D_s$ decays \cite{MEL} favouring a mixing
angle in the range $-18^\circ \leq \theta_P \leq -10^\circ$ with the best
agreement observed for $\theta_P = -14^\circ$. Similarly, from the measurement
of the $\pi^+\pi^-$ invariant-mass distribution
in $\eta\prime\rightarrow\pi^+\pi^-\gamma$
\cite{ABE} one can deduce, depending on the model, either
$\theta_{P}=-16.44^\circ\pm 1.20^\circ$ $(\varphi_{P}=38.30^\circ\pm
1.20^\circ)$ or
$\theta_{P}=-23.24^\circ\pm 1.23^\circ$ $(\varphi_{P}=31.50^\circ\pm
1.23^\circ)$ while a recent analysis of the $\eta$ and $\eta\prime$
radiative decays into $\gamma l^+l^-$ and $\gamma \pi^+\pi^-$ 
leads to $\theta_P \sim -16.5$ \cite{IVA}.
 Finally, from the study of the 
photon-meson transition form factors \cite{ANI} a value of 
$\theta_{P}=-16.7^\circ\pm 2.8^\circ$ has been determined.

One can safely conclude this miscellaneous section saying that $\theta_{P} 
\simeq -16^\circ$ is favoured by all these recent and independent results.
\section{Conclusions}
We have made a rather exhaustive analysis of the pseudoscalar
$\eta$-$\eta\prime$ mixing angle using well established and accepted
phenomenology and the experimental data available at present. We have
surveyed various types of data and found that the strong decays of 
tensor-mesons $T(2^{++})\rightarrow PP$ and higher-spin mesons 
$M_J\rightarrow PP$, for which unfortunately one cannot account for 
SU(3)-breaking corrections, favour the choice of 
$\theta_{P} \simeq -13^\circ$; essentially the same value is also favoured by 
the two-photon annihilation decays $P\rightarrow\gamma\gamma$.
Other data such as the radiative decays
$V\rightarrow P\gamma$ and $P\rightarrow V\gamma$, $J/\psi$ decays into a
vector and a pseudoscalar, together with other types of transitions favour
the choice of $\theta_P\simeq -17^\circ$. We should emphasize that our 
conclusions are based on the assumptions of the simple $\eta$-$\eta\prime$ 
mixing scenario, the use of the $SU(3)$ and nonet symmetry and the manner in 
which $SU(3)$-breaking corrections are introduced. 
In summary, we have just shown that present data
are consistent with a mixing angle in the range of $\theta_P\simeq -17^\circ$
and $\theta_P\simeq -13^\circ$. A weighted average value of
$\theta_{P}=-15.5^\circ\pm 1.3^\circ$ seems to be favoured by the different
types of decays involved in the analysis.

\clearpage
\begin{table}
\caption{Strong decays of spin-two, tensor mesons into pseudoscalar pairs,
$T(2^{++})\rightarrow PP$. The three columns display the various decay
modes, the corresponding coupling constants and the experimental branching 
ratios (BR) from Ref.~\protect\cite{PDG}, respectively. 
Consistent values for the mixing
angle $\varphi_P$ ($\theta_P\simeq \varphi_P-54.7^\circ$) are obtained by 
fitting the BR's of each separate tensor meson. Values for $\varphi_T$ are 
simultaneously predicted.}
\vspace*{0.4cm}
\centering
\begin{tabular}{ccc}\hline\hline\\
decay mode & $g_{TPP}/2g$ & BR(\%) \\[1ex]
& & mixing angle(s) \\[1ex] \hline\\
$a_2\rightarrow K\bar K$ & $1$ &   $4.9\pm 0.8$ \\[1ex]
$a_2\rightarrow \eta\pi$ & $\sqrt{2}\cos\varphi_{P}$ & $14.5\pm 1.2$ \\[1ex]
$a_2\rightarrow \eta\prime\pi$ & $\sqrt{2}\sin\varphi_{P}$ & $0.57\pm 0.11$
\\[1ex] \hline\\
& & $\varphi_{P}=43.2^\circ\pm 2.8^\circ$ \\[1ex] \hline\\
$K^\ast_2\rightarrow K\pi$ & $\sqrt{3}/\sqrt{2}$ & $49.7\pm 1.2$ \\[1ex]
$K^\ast_2\rightarrow K\eta$ & 
$\frac{1}{\sqrt{2}}\cos\varphi_{P}-\sin\varphi_{P}$ &
$0.14^{+0.28}_{-0.09}$ \\[1ex] \hline\\
& & $\varphi_{P}=40.7^\circ\pm 3.7^\circ$ \\[1ex] \hline\\
$f\rightarrow \pi\pi$ & $\sqrt{3}\cos\varphi_{T}$ & $84.7^{+2.6}_{-1.2}$
\\[1ex]
$f\rightarrow K\bar K$ & $\cos\varphi_{T}-\sqrt{2}\sin\varphi_{T}$ &
$4.6\pm 0.5$ \\[1ex]
$f\rightarrow \eta\eta$ & $\cos\varphi_{T}\cos^{2}\varphi_{P}-
\sqrt{2}\sin\varphi_{T}\sin^{2}\varphi_{P}$ & $0.45\pm 0.10$ \\[1ex] 
\hline\\
& & $\varphi_{P}=42.7^\circ\pm 5.4^\circ$ \\[1ex]
& & $\varphi_{T}=-7.8^\circ\pm 2.6^\circ$ \\[1ex] \hline\\
$f\prime\rightarrow \pi\pi$ & $\sqrt{3}\sin\varphi_{T}$ &  $0.82\pm 0.15$
\\[1ex]
$f\prime\rightarrow K\bar K$ & $\sin\varphi_{T}+\sqrt{2}\cos\varphi_{T}$ &
$88.8\pm 3.1$ \\[1ex]
$f\prime\rightarrow \eta\eta$ & $\sin\varphi_{T}\cos^{2}\varphi_{P}+
\sqrt{2}\cos\varphi_{T}\sin^{2}\varphi_{P}$ & $10.3\pm 3.1$ \\[1ex] \hline\\
& & $\varphi_{P}=41.0^\circ\pm 3.5^\circ$ \\[1ex]
& & $\varphi_{T}=-2.3^\circ\pm 0.2^\circ$\\[1ex] \hline\hline
\end{tabular}
\label{tableTPP}
\end{table}
\begin{table}
\caption{Strong decays of spin-three and spin-four mesons into pseudoscalar
pairs, $M_J\rightarrow PP$. As in Table I, a value of the mixing angle
$\varphi_P$ is obtained from each set of BR's. The fit gives also a value for 
$\varphi_4$, the mixing angle in the spin-four nonet.}
\vspace*{0.4cm}
\centering
\begin{tabular}{ccc}\hline\hline\\
decay mode & $g_{M_JPP}/2g$ & BR(\%) \\[1ex]
& & mixing angle(s) \\[1ex] \hline\\
$f_4\rightarrow \pi\pi$ & $\sqrt{3}\cos\varphi_{4}$ & $17.0\pm 1.5$ \\[1ex]
$f_4\rightarrow K\bar K$ & $\cos\varphi_{4}-\sqrt{2}\sin\varphi_{4}$ &
$0.68^{+0.34}_{-0.18}$ \\[1ex]
$f_4\rightarrow \eta\eta$ & $\cos\varphi_{4}\cos^{2}\varphi_{P}-
\sqrt{2}\sin\varphi_{4}\sin^{2}\varphi_{P}$ & $0.21\pm 0.08$ \\[1ex] 
\hline\\
& & $\varphi_{P}=41.2^\circ\pm 3.7^\circ$ \\[1ex]
& & $\varphi_{4}=15.7^\circ\pm 4.4^\circ$ \\[1ex] \hline\\
$K_3^\ast\rightarrow K\pi$ & $\sqrt{3}/\sqrt{2}$ & $19.3\pm 1.0$ \\[1ex]
$K_3^\ast\rightarrow K\eta$ &
$\frac{1}{\sqrt{2}}\cos\varphi_{P}+\sin\varphi_{P}$ &
$8.0\pm 1.5$ \\[1ex] \hline\\
& & $\varphi_{P}=50^\circ\pm 26^\circ$\\[1ex] \hline\hline
\end{tabular}
\label{tableMJPP}
\end{table}
\begin{table}
\caption{Radiative decays of light mesons, $V\rightarrow P\gamma$ and
$P\rightarrow V\gamma$. Columns are organized as in the preceding Tables, 
but here $SU(3)$-breaking corrections are introduced in terms of
constituent quark mass differences $\bar m/m_s\simeq 1/1.45$. 
The small mixing angle $\varphi_V$ signalling departure of $\omega$ 
and $\phi$ from ideal mixing is not neglected and left as a free 
parameter in the fit. The resulting values for $\varphi_P$ and $\varphi_V$ 
are displayed. The value of the full widths used in the fit are: 
$\Gamma_\rho=150.7\pm 1.2$ MeV, $\Gamma_\omega=8.43\pm 0.10$ MeV,
$\Gamma_\phi=4.43\pm 0.05$ MeV and $\Gamma_{\eta\prime}=0.201\pm 0.016$ MeV.}
\vspace*{0.4cm}
\centering
\begin{tabular}{ccc}\hline\hline\\
decay mode & $g_{VP\gamma}/g$ & BR(\%) \\[1ex]
& & mixing angle(s) \\[1ex] \hline\\
$\rho^{0}\rightarrow \eta\gamma$ & $\cos\varphi_{P}$ &   $(3.8\pm 0.7)\
10^{-2}$ \\[1ex]
$\rho^{0}\rightarrow \pi^{0}\gamma$ & $1/3$ & $(7.9\pm 2.0)\ 10^{-2}$ \\[1ex]
$\rho^{\pm}\rightarrow \pi^{\pm}\gamma$ & $1/3$ & $(4.5\pm 0.5)\ 10^{-2}$
\\[1ex]
$\omega\rightarrow \eta\gamma$ &
$\frac{1}{3}(\cos\varphi_{P}\cos\varphi_{V}-2
\frac{\bar m}{m_s}\sin\varphi_{P}\sin\varphi_{V})$ &
$(8.3\pm 2.1)\ 10^{-2}$ \\[1ex]
$\omega\rightarrow \pi^{0}\gamma$ & $\cos\varphi_{V}$ & $8.5\pm 0.5$ \\[1ex]
$\phi\rightarrow \eta\gamma$ &
$\frac{1}{3}(\cos\varphi_{P}\sin\varphi_{V}+2
\frac{\bar m}{m_s}\sin\varphi_{P}\cos\varphi_{V})$ & $1.26\pm 0.06$ \\[1ex]
$\phi\rightarrow \eta\prime\gamma$ &
$\frac{1}{3}(\sin\varphi_{P}\sin\varphi_{V}-2
\frac{\bar m}{m_s}\cos\varphi_{P}\cos\varphi_{V})$ &
$<4.1\ 10^{-2}$ CL=90\% \\[1ex]
$\phi\rightarrow \pi^{0}\gamma$ & $\sin\varphi_{V}$ & 
$(1.31\pm 0.13)\ 10^{-1}$\\[1ex]
$\eta\prime\rightarrow \rho\gamma$ & $\sin\varphi_{P}$ &  
$30.2\pm 1.3$ \\[1ex]
$\eta\prime\rightarrow \omega\gamma$ &
$\frac{1}{3}(\sin\varphi_{P}\cos\varphi_{V}+2
\frac{\bar m}{m_s}\cos\varphi_{P}\sin\varphi_{V})$ &
$3.02\pm 0.30$ \\[1ex] \hline\\
& & $\varphi_{P}=36.5^\circ\pm 1.4^\circ$ \\[1ex]
& & $\varphi_{V}=3.4^\circ\pm 0.2^\circ$\\[1ex] \hline\hline
\end{tabular}
\label{tableVPg&PVg}
\end{table}
\begin{table}
\caption{Two-photon annihilation decays 
$\pi^0$, $\eta$, $\eta\prime\rightarrow\gamma\gamma$. 
As in the previous Table, SU(3)-breaking effects are introduced and a new 
value for $\varphi_P$ is obtained.}
\vspace*{0.4cm}
\centering
\begin{tabular}{ccc}\hline\hline\\
decay mode & $g_{P^0\gamma\gamma}/g$ & Decay width \\[1ex]
& & mixing angle \\[1ex] \hline\\
$\pi^{0}\rightarrow \gamma\gamma$ & $\frac{1}{3\sqrt{2}}$ 
& $7.74\pm 0.55$ eV 
\\[1ex]
$\eta\rightarrow \gamma\gamma$ &
$\frac{5}{9\sqrt{2}}(\cos\varphi_{P}-\frac{\sqrt{2}}{5}
\frac{\bar m}{m_s}\sin\varphi_{P})$ &
$0.46\pm 0.04$ keV\\[1ex]
$\eta\prime\rightarrow \gamma\gamma$ &
$\frac{5}{9\sqrt{2}}(\sin\varphi_{P}+\frac{\sqrt{2}}{5}
\frac{\bar m}{m_s}\cos\varphi_{P})$ &
$4.26\pm 0.43$ keV\\[1ex] \hline\\
& & $\varphi_{P}=42.4^\circ\pm 2.0^\circ$\\[1ex] \hline\hline
\end{tabular}
\label{tablePgg}
\end{table}
\begin{table}
\caption{$J/\psi$ decays into a vector and a pseudoscalar meson,
$J/\psi\rightarrow VP$. A value of $\varphi_P$ deduced from a partial fit
including isospin $I=1$ final states is shown. A detailed description of the
parameters involved in the amplitudes, and details about the fit can be 
found in Ref.~\protect\cite{BES}.}
\vspace*{0.4cm}
\centering
\begin{tabular}{ccc}\hline\hline\\
decay mode & $g_{J/\psi VP}$ & BR($10^{-3}$) \\[1ex] 
& & mixing angle \\[1ex] \hline\\
$J/\psi\rightarrow \rho\eta$ & $3e\cos\varphi_{P}$ &   
$0.193\pm 0.023$ \\[1ex]
$J/\psi\rightarrow \rho\eta\prime$ & $3e\sin\varphi_{P}$ & $0.105\pm 0.018$
\\[1ex]
$J/\psi\rightarrow \omega_{NS}\pi^0$ & $3e$ & $0.42\pm 0.06$ \\[1ex]
\hline\\
& & $\varphi_{P}=40.2^\circ\pm 2.8^\circ$ \\[1ex] \hline\\
$J/\psi\rightarrow \rho\pi$ & $g + e$ & $12.8\pm 1.0$ \\[1ex]
$J/\psi\rightarrow K^{\ast\pm}K^{\ast\mp}$ & $g(1-s) + e(2-x)$ &
$5.0\pm 0.4$ \\[1ex]
$J/\psi\rightarrow  K^{\ast 0}K^{\ast 0}$ & $g(1-s) - 2e(1+x)/2$ & 
$4.2\pm 0.4$\\[1ex]
$J/\psi\rightarrow \omega_{NS}\eta$ & $(g+e) X_{\eta} +
\sqrt2 r g (\sqrt2 X_{\eta}+Y_{\eta})$ &
$1.58\pm 0.16$ \\[1ex]
$J/\psi\rightarrow \omega_{NS}\eta\prime$ & $(g+e) X_{\eta\prime} +
\sqrt2 rg (\sqrt2 X_{\eta\prime}+Y_{\eta\prime})$ &
$0.167\pm 0.025$ \\[1ex]
$J/\psi\rightarrow \phi_S\eta$ & $[g(1-2s)-2ex] Y_{\eta} +
r g(1-s)(\sqrt2X_{\eta}+Y_{\eta})$ &  $0.65\pm 0.07$ \\[1ex]
$J/\psi\rightarrow \phi_S\eta\prime$ & $[g(1-2s)-2ex] Y_{\eta\prime} +
r g(1-s)(\sqrt2X_{\eta\prime}+Y_{\eta\prime})$ &
$0.33\pm 0.04$ \\[1ex]
$J/\psi\rightarrow \phi_S\pi^0$ & $0$ & $<0.0068$ CL=90\% \\[1ex] \hline\hline
\end{tabular}
\label{tableJVP}
\end{table}
\end{document}